# Characterization of atmospheric coherent structures and their impact on a utility-scale wind turbine

Aliza Abraham[1,2] and Jiarong Hong[1,2,*]

[1]St. Anthony Falls Laboratory, University of Minnesota, 2 Third Avenue SE, Minneapolis, MN 55455, USA

[2]Department of Mechanical Engineering, University of Minnesota, 111 Church Street SE, Minneapolis, MN 55455, USA

*Corresponding author: jhong@umn.edu

Abstract

Atmospheric turbulent velocity fluctuations are known to increase wind turbine structural loading and accelerate wake recovery, but the impact of vortical coherent structures in the atmosphere on wind turbines has not yet been evaluated. The current study uses flow imaging with natural snowfall with a field of view spanning the inflow and near wake. Vortical coherent structures with diameters on the order of 1 m are identified and characterized in the flow approaching a 2.5 MW wind turbine in the region spanning the bottom blade tip elevation to hub height. Their impact on turbine structural loading, power generation, and wake behaviour are evaluated. Long coherent structure packets ($\gtrsim$ 200 m) are shown to increase fluctuating stresses on the turbine support tower. Large inflow vortices interact with the turbine blades, leading to deviations from the expected power generation. The sign of these deviations is related to the rotation direction of the vortices, with rotation in the same direction as the circulation on the blades leading to periods of power surplus, and opposite rotation causing power deficit. Periods of power deficit coincide with wake contraction events. These findings highlight the importance of considering coherent structure properties when making turbine design and siting decisions.

Impact Statement
Utility-scale wind turbines are subject to complex atmospheric conditions, including varying levels of turbulence. Atmospheric turbulence is characterized in part by coherent vortical structures, i.e., strongly rotating regions of fluid, though they are difficult to study in real atmospheric flow due to limitations in measurement techinques. The current study characterizes these coherent structures using a super-large-scale flow visualization technique with natural snowfall, and shows that they lead to increased stress on the turbine structure, which can cause premature structural failure. This study also shows that coherent structures impact power generation and the behaviour of the wake (the region of slower air behind the turbine), which affects overall wind farm efficiency. By quantifying the characteristics of these structures and their impact on wind turbines, these findings highlight the importance of including considerations of coherent structure properties in wind farm design and siting decisions.

# 1. Introduction

Wind turbines operate in a complex atmospheric environment subject to constantly changing turbulent wind conditions that are influenced by a broad range of scales (Veers et al., 2019). Previous works have established the significant impacts of atmospheric turbulent velocity fluctuations on wind turbines, showing that higher levels of turbulence intensity lead to increased structural loading (Frandsen, 2007; Park, Basu, & Manuel, 2014) and accelerated recovery of the wake, the region of slower air behind the turbine (Medici & Alfredsson, 2006; Wu & Porte-Agel, 2012). Moreover, turbulent wind speed fluctuations lead to power production deviations from the expected power curve, contributing to uncertainty in wind farm energy production and increased financial risk for energy contractors (Lee & Fields, 2021). However, the impact of atmospheric coherent structures, characteristic features of turbulence, has not been thoroughly explored.

Coherent structures in the atmospheric boundary layer (ABL) have been observed to exhibit similar properties to those of canonical turbulent boundary layers observed in laboratory experiments. In particular, signatures of hairpin vortex packets have been consistently identified in the ABL within ~10-20 m elevations (Hommema & Adrian, 2003; Li & Bou-Zeid, 2011; Oncley, Hartogensisa, & Tong, 2016; Heisel et al., 2018a). These structures are more difficult to detect at higher elevations where they could interact with wind turbines due to limitations in the spatial resolution of conventional measurement techniques such as sonic anemometers. However, some recent studies have used lidar to detect intermittent coherent structures at elevations up to 200 m. Träumner, et al. (2015) observed large-scale coherent regions of velocity, or "streaks", with 100-200 m vertical extent that are interpreted as indicators of groups of hairpin vortices, with insufficient resolution to detect individual vortex cores. By qualitatively classifying hourly intervals as "with structures" or "homogeneous", they determined that the mean wind speed is higher for periods with structures than for homogeneous periods (4.5-6.3 m/s vs. 2.2-3.8 m/s, respectively). Cheliotis et al. (2020) identified similar velocity streaks in lidar data at an elevation of 75 m. Using an automated classification method, they detected coherent streaks on 25% of the lidar scans over a period of 2 months. These streaks were found to occur more frequently at night than during the day. Alcayaga et al. (2020) directly quantified the vertical component of vorticity from lidar scans taken at 200 m of elevation and found that positive-divergence streaks (indicative of vertical ejections of low-momentum flow) are bounded by counter-rotating vortices. These features are characteristic of canonical hairpin vortex packets found in turbulent boundary layers at a wide range of Reynolds numbers, which have been shown to extend beyond the logarithmic layer even up to the edge of the boundary layer (Adrian, 2007).

Very few works have investigated the impact of coherent structures on wind turbines. One such study simulated a Kelvin Helmholtz billow, which generates coherent structures in the atmosphere, interacting with a wind turbine. Using large eddy simulations supplemented with a short example from field data, they found that coherent structures cause high frequency structural vibrations and strong impulsive loading events, both of which can lead to structural damage (Kelley et al. 2005). A more recent investigation used cylinders of different sizes to generate turbulence upstream of a hydrokinetic turbine in open-channel flow. This study showed that cylinders placed farther upstream cause the vortex structures in the turbine wake to breakdown faster than those placed closer to the turbine, because the upstream placement allows longer time for coherent structures to develop. In addition, the slope of the turbine power spectrum is steeper when in the wake of a cylinder in the same intermediate frequency range that scales with the cylinder diameter and flow speed due to von Kármán vortex shedding (Chamorro et al., 2015a). Though these studies provided important insights into the interactions between wind turbines and coherent structures, they were not conducted under an extended period of real atmospheric conditions.

Previous works have had limited success directly characterizing atmospheric coherent structures over a range of scales and determining their impact on wind turbines due to limitations in spatio-temporal resolution. Of the studies discussed above, only Alcayaga et al. (2020) directly identified individual vortices at high enough elevations to interact with wind turbines, and even they were constrained by their use of scanning lidar, which requires 45 seconds to obtain a full scan of the measurement region (Karagali et al. 2018). On the other hand, several field-scale investigations have explored the interaction between wind turbines and different ranges of atmospheric turbulent length scales. Chamorro et al. (2015b) conducted a field study using a meteorological tower to characterize the inflow experienced by the 2.5 MW wind turbine at the Eolos Wind Energy Research Station, and investigated the modulation of power generation and foundation strain. They identified three frequency regions that influence power and strain differently: subrotor length scales have no effect on power but directly influence strain, power and strain both exhibit a damped response to intermediate length scales, and the largest length scales (on the order of the boundary layer thickness) directly influence strain and power. Heisel, Hong, & Guala (2018b) quantified the modulation of turbulent length scales between the inflow, probed using a meteorological tower, and the wake, measured with lidar. They observed a reduction of low-frequency energy and an increase in high-frequency energy in the wake relative to the inflow, suggesting a breakdown of large turbulent scales into smaller scales. This sheltering effect was only observed when the turbine was operating in the optimal regime. Both aforementioned studies used frequency spectra to analyse the turbulent flow scales, but they did not directly detect coherent structures. However, frequency spectra do not fully capture the complexity of the ABL, including the coherent structures discussed above.

To summarize, direct characterization of atmospheric coherent structures has been hindered by the low spatio-temporal resolution of conventional field-scale measurement techniques. Furthermore, investigations into the impact of such structures on wind turbines have been limited to the laboratory and simulations where the stochasticity of atmospheric flow is difficult to replicate. Therefore, the goal of the current investigation is to 1) provide a more detailed characterization of atmospheric coherent structures, including their intermittency and stochasticity, and 2) quantify the impact of these structures on utility-scale wind turbine loading, power generation, and wake behaviour under real atmospheric conditions. These aims are achieved using snow-powered super-large-scale particle image velocimetry, a high-resolution field scale measurement technique that has proven successful in quantifying flows (including coherent vortical structures) around utility-scale wind turbines (Hong & Abraham, 2020). Previous studies have used this method to characterize the incoming flow approaching the 2.5 MW turbine at the Eolos site (Li et al. 2020) and to identify previously unobserved wake behaviours including wake contraction in response to changing turbine blade pitch (Dasari et al. 2019). This technique has also been used to characterize the ABL flow at the Eolos site (Toloui et al., 2014) and to provide insight into the structure of very high Reynolds number turbulent boundary layers (Heisel et al., 2018). In the current study, we extend this technique to identify coherent structures in the inflow and to quantify the turbine and wake response. The paper is structured as follows: Section 2 describes the experimental methods, Section 3 presents the results of the investigation, and Section 4 provides a summary and discussion of the key findings.

## 2. Methodology

### 2.1. Experimental setup

The field experiments were conducted at the University of Minnesota Eolos Wind Energy Research Field Station in Rosemount, Minnesota. The site hosts a 2.5 MW Clipper Liberty C96 wind turbine with a hub height of 80 m and a rotor diameter ($D$) of 96 m. The turbine is a horizontal axis, three-bladed, pitch regulated machine, fully instrumented with a supervisory control and data acquisition (SCADA) system at the nacelle and 20 strain gauges around the base of the support tower. The SCADA data is recorded at 20 Hz and includes atmospheric conditions such as wind speed and direction collected by a sonic anemometer at the back of the nacelle, and turbine operational parameters including blade pitch, rotor speed, nacelle orientation, and power generation. A meteorological tower (met tower) is located 170 m south of the turbine, with sensors at 7 m, 27 m, 52 m, 77 m, 102 m, and 126 m elevations recording wind speed and direction (cup and vane anemometers), temperature, and relative humidity at 1 Hz. Additional sonic anemometers at 10 m, 30 m, 80 m, and 129 m record three wind speed components and temperature at 20 Hz. The area surrounding the turbine is primarily flat farmland with sparse shrubs located 100 m upstream, beyond which lies 1 km of additional farmland. No structures or large trees are located in the immediate upstream vicinity (i.e., within $10D$ upstream) of the turbine. The ground was covered with snow during all data collection periods discussed in the current study. Further details about the site and turbine can be found in Hong et al. (2014) and Dasari et al. (2019).

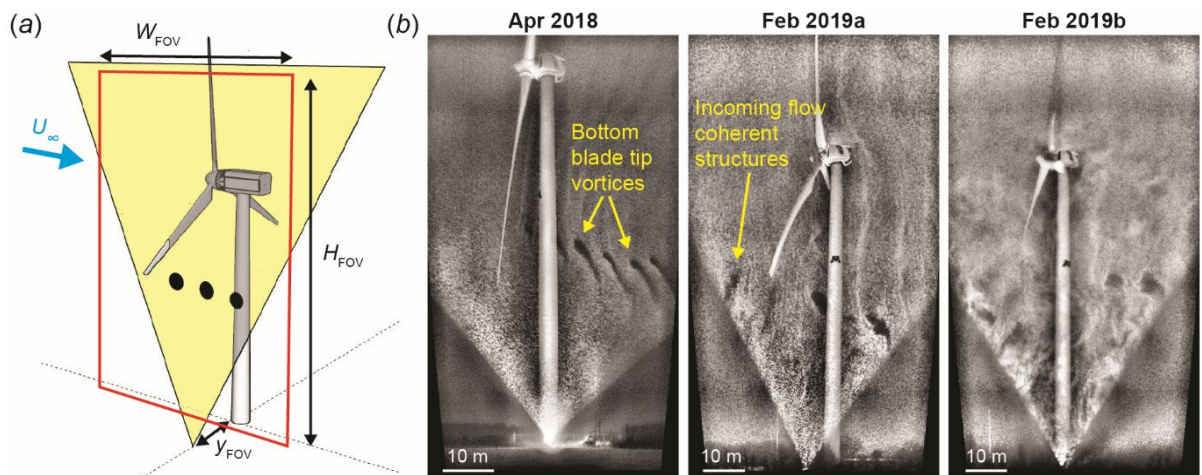

Fig. 1. (a) Schematic showing the flow visualization FOV for the experiments. (b) Sample enhanced and de-warped flow visualization images for each of the three datasets. Arrows indicate vortices shed from the bottom blade tips in the near wake and coherent structures in the inflow, both seen as voids in the snow images.

In addition to the data collected by the turbine, flow visualization data is recorded using super-large-scale particle image velocimetry (SLPIV), with natural snowflakes serving as flow tracers (Hong & Abraham, 2020). A collimated searchlight is projected into a sheet to illuminate the snowflakes in the field of view (FOV), which spans the inflow and near wake of the turbine (Figure 1a). The light sheet is slightly offset from the central tower plane in the spanwise direction ($y_{FOV}$ in Table 1) to capture the vortices shed from the turbine blade tips, which mark the wake boundary, without distortion from the turbine support tower. A Nikon D600 camera, fitted with a 50 mm $f$/1.2 lens, is used to record videos of the snowflake motion at 30 Hz. Three flow visualization datasets were recorded during snowstorms that occurred during the hours of darkness, one on April 9, 2018, and two on February 24, 2019, two hours apart (Figure 1b). The conditions of each dataset vary in mean wind speed at hub height ($U_\infty$), region of turbine operation, and atmospheric conditions (Table 1). The atmospheric stability of each dataset is quantified using the Bulk Richardson Number, $R_B = \frac{g \Delta \overline{\theta_v} \Delta z}{\overline{\theta_v}[(\Delta \overline{U})^2 + (\Delta \overline{V})^2]}$, where $g$ is the gravitational acceleration, $\overline{\theta_v}$ is the virtual potential temperature, $z$ is the elevation, $U$ is the northerly wind component, and $V$ is the westerly wind component (Stull, 1988). All parameters are calculated using the cup and vane anemometers at the 126 m and 7 m elevations of the met tower. We also calculate the Monin-Obukhov length for each dataset as $L_{OB} = -U_\tau^3 \overline{\theta_v}/\kappa g \overline{w'\theta_v'}$, where $U_\tau$ is the friction velocity, $\kappa$ is the von Kármán constant, and $\overline{w'\theta_v'}$ is the vertical heat flux calculated using temperature and vertical wind speed fluctuations from the 10 m sonic anemometer on the met tower. The friction velocity is estimated using the Reynolds stresses (Stull, 1988), as follows: $U_\tau = (\overline{u'w'}^2 + \overline{v'w'}^2)^{1/4}$, where $u'$, $v'$, and $w'$ are the fluctuating streamwise, spanwise, and vertical velocity components, respectively, also measured at the 10 m sonic anemometer. The ABL during both Feb 2019 datasets can be considered neutral, as $R_B$ is less than the critical value of 0.25 but greater than 0, and $z/L_{OB} \ll 0.1$. The Apr 2018 dataset is slightly stable. The ABL is typically near-neutral during snow storms, but the snow was relatively weak on the night the Apr 2018 data was collected, leading to a shift towards the more stable stratification typical of night-time. Though capturing approximately the same area, the dimensions of the FOVs ($H_{FOV} \times W_{FOV}$) for each dataset vary slightly as well. The Apr 2018 dataset is slightly shorter than the other two, though the turbine is producing power for a substantially lower percentage of the time. Therefore, this dataset is primarily used for inflow characterization and contributes very little to the analysis of coherent structure impact on turbine operation.

Table 1. Parameters for each of the three datasets

| | Apr 2018 | Feb 2019a | Feb 2019b |
|---|---|---|---|
| Data collection date and time (UTC) | 04:35:39 Apr. 9, 2018 | 04:34:00 Feb. 24, 2019 | 06:39:08 Feb. 24, 2019 |
| $U_\infty$ (m/s) | 2.6 | 7.0 | 11.4 |
| Mean wind direction (° clockwise from North) | 59.3 | 336.6 | 322.6 |
| Region of turbine operation | 1-1.5 | 1.5-2 | 2.5-3 |
| $R_B$ | 0.45 | 0.12 | 0.04 |
| $L_{OB}$ (m) | 35 | 1550 | -2260 |
| $U_\tau$ (m/s) | 0.07 | 0.41 | 0.73 |
| $H_{FOV} \times W_{FOV}$ (m × m) | 87 × 49 | 105 × 59 | 93 × 52 |
| $y_{FOV}$ (m) | 19 | 21 | 23 |
| Duration | 18 min 30 sec | 20 min | 20 min |
| Percentage of time turbine is producing power | 3% | 100% | 100% |

## 2.2. Inflow coherent structure identification

In the flow visualization images, regions of strong vorticity are visible as dark voids where the snowflake concentration is significantly reduced (see Figure 1b). These voids are caused by the rotating fluid expelling snowflakes from the centre of the vortex. In previous studies, these voids have been used to characterize the behaviour of vortices shed from the tips of the turbine blades (Hong et al. 2016, Dasari et al. 2019, Abraham & Hong 2020, Abraham & Hong 2021). In the current study, we also analyse these voids to characterize the level of coherent vortical structures in the inflow approaching the turbine. There are many types of coherent structures in a turbulent boundary layer, but here we focus on coherent structures that leave signatures of vortices on our measurement plane such as the hairpin vortex packets reported in the literature (Hommema & Adrian, 2003; D. Li & Bou-Zeid, 2011; Oncley et al., 2016; Heisel et al., 2018a; Alcayaga et al., 2020), as mentioned in the introduction. Note that the voids visible in the snow visualization images only capture a cross-section of the coherent structures as determined by the illuminated plane of the light sheet, so their three-dimensional structure (e.g., if they are true hairpin vortices) cannot be characterized. However, many previous studies conducted at a range of Reynolds numbers have successfully utilized two-dimensional datasets to analyse three-dimensional structures (e.g., Adrian, Meinhart, & Tomkins, 2000; Heisel et al., 2018a; Wu & Christensen, 2006).

First, the images are enhanced using wavelet denoising and adaptive histogram equalization to strengthen the void signature over the background noise. Next, they are de-warped to correct for distortion caused by the inclination angle of the camera relative to the ground (Dasari et al., 2019; Toloui et al., 2014). From the enhanced and de-warped images, a region located 15 m upstream of the turbine and ranging from 35 m to 80 m in elevation is selected, angled such that the turbine blades are excluded (Figure 2). In some video frames, coherent structures are clearly seen within this window (Figure 2a), whereas no coherent structures are visible in others (Figure 2b). MATLAB image category classification using "bag of features" (also known as "bag of words", MathWorks, 2020) is used to classify the video frames from all three datasets as coherent (labelled as 1) or non-coherent (labelled as 0). To train the classifier, 758 frames taken from all three datasets were manually identified as containing coherent structures (371 frames) or not containing coherent structures (387 frames) based on the presence or absence of dark voids in the images (Figure 2c shows a gallery of these images). Of the 758 manually classified frames, 60% were randomly selected for training and the other 40% were reserved for validation. Features were then extracted from all training images and a "visual vocabulary" of a reduced number of feature clusters is defined using K-means clustering. The images were categorized based on the frequency of occurrence of each feature cluster in the image. Next, these feature clusters were used to classify the validation images and the results were compared with the manual classification. The automatic classifier yielded an average accuracy of 90%, with 89% of non-coherent frames accurately identified and 90% of coherent frames accurately identified. The trained classifier was then used to categorize all 105,189 video frames from the three datasets. A manual check was performed on several segments of automatically classified frames to ensure the accuracy of the classifier. It is worth noting that previous studies (i.e., Cheliotis et al., 2020; Träumner et al., 2015) have used manual visual identification to classify periods as with or without coherent structures, and Cheliotis et al. (2020) also used this manual classification method to train a machine learning-based classifier. Such an automated method eliminates the need for manually selected filtering or thresholding parameters. With every frame of all three datasets classified, periods with a strong presence of coherent structures can be compared to those without structures in the inflow.

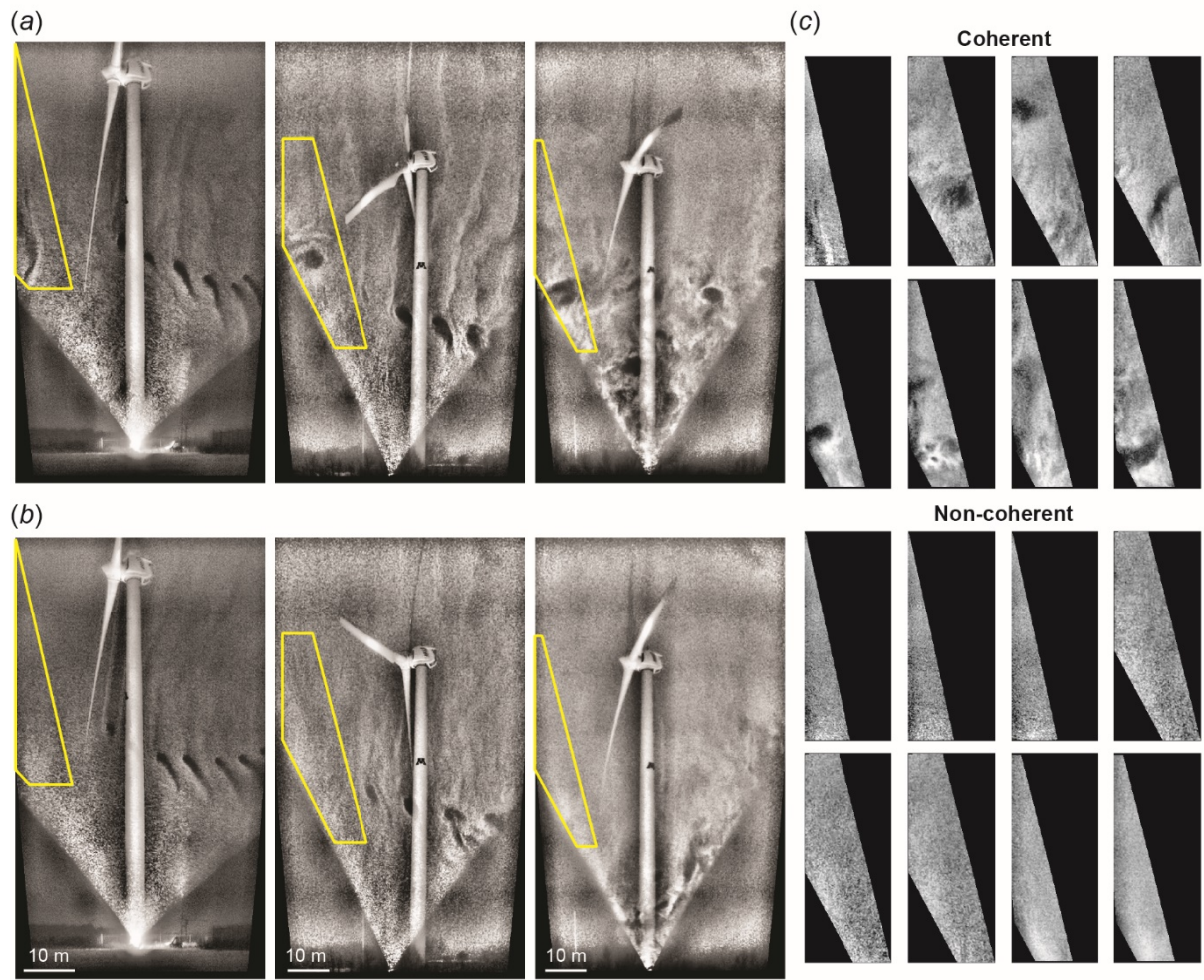

Fig. 2. Sample frames for each of the three datasets (a) with coherent structures and (b) without coherent structures in the inflow. The yellow outline indicates the region used for the machine learning classification. Note that the physical locations of these sampling regions are the same for all three datasets, though they appear different due to the differences in the FOVs. (c) A gallery of example coherent and non-coherent images used to train the classifier.

## 2.3. Particle image velocimetry with blade skipping

Particle image velocimetry (PIV) processing is conducted using PIVlab, an open-source MATLAB-based PIV software (Thielicke & Stamhuis, 2014). As with previous studies investigating such a large FOV, the snowflake patterns in the enhanced images provide the signal for the PIV correlation rather than individual snowflakes (Abraham, Dasari, & Hong, 2019; Dasari et al., 2019; Abraham & Hong, 2021). In the current study, the velocity field is calculated over the inflow region, spanning from 16.4 m to 82.5 m in elevation, and from 31.0 m to 9.5 m upstream. An interrogation window of 64 × 64 pixels with 50% overlap is used for the first pass, and a second pass is conducted with an interrogation window of 48 × 48 pixels with 50% overlap. These window sizes yield a spatial resolution of 2.7 m for the velocity vector field.

The PIV code is modified to account for the turbine blades passing through the analysed FOV, as the motion of the blades would interfere with the flow field calculation. First, the blades are detected in each frame by computing the correlation between all images in a sequence. As the blades are the strongest feature in the images, the frames with the highest correlation have the blades in the same location. These frames are averaged and converted to binary images that only contain masks of the blades, which are used to determine if a blade is within a PIV interrogation window. For each interrogation window, the frames with a blade detected are skipped during the PIV correlation step (Figure 3a), and the velocity vector at that location is calculated by correlating the frames before and after the blade passes through (Figure 3b). The magnitude of this vector is divided by the number of frames skipped to account for the additional displacement that occurred during those frames. Note that the snowflake patterns persist with a clear enough signature to yield a strong correlation despite the skipped frames (Figure 3c). Additionally, only the windows through which the blade passes undergo frame skipping, while the remaining parts of the image are unaffected. Still, the temporal resolution of the flow velocity calculated when the blade is passing through the window is limited by the speed of the blade at each radial location. The slowest blade speed occurs near the blade root, where the blade remains within a window for a maximum of 25 frames. In this worst-case-scenario, the temporal resolution of the velocity data is 0.8 seconds in the region near the blade root, compared to the upper limit of 0.03 seconds as determined by the camera frame rate.

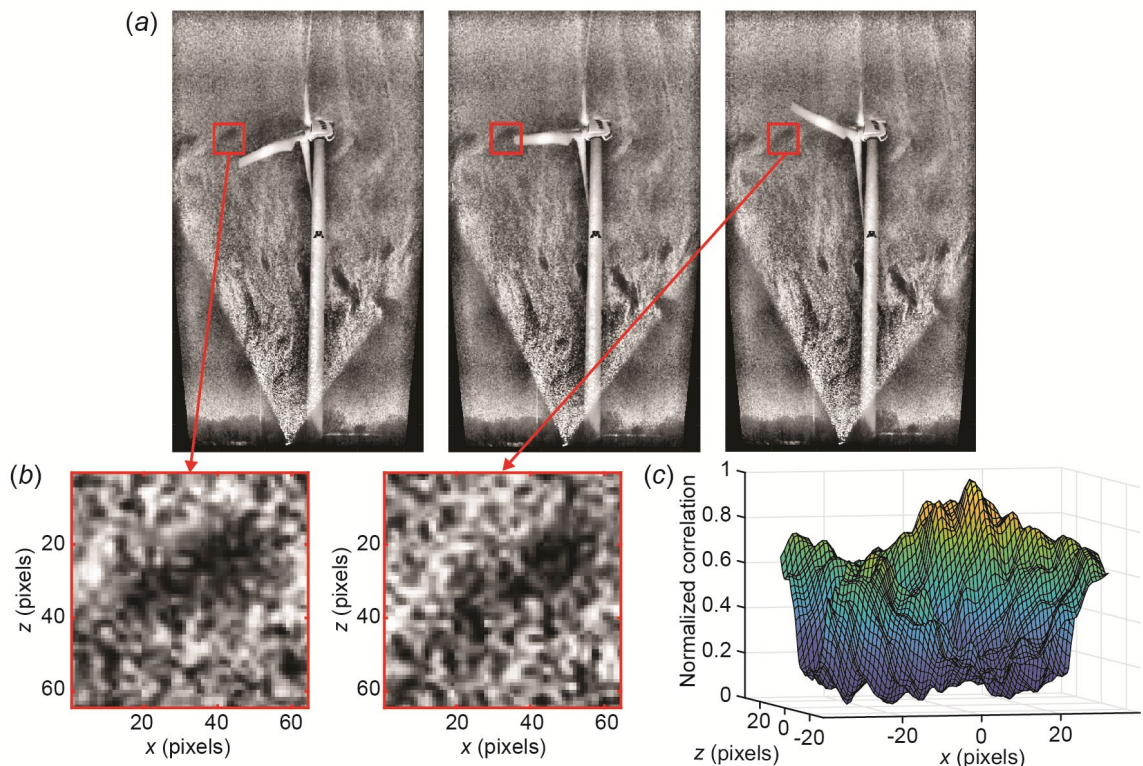

Fig. 3. Demonstration of the PIV blade-skipping algorithm, including (a) sample images with one PIV window (64 × 64 pixels) outlined in red. The first image is the frame before the blade enters the window, the second is a frame with the blade inside the window, and the third is the frame just after the blade moves outside the window. The windows from the first and third images, exhibiting a clear pattern persisting across the frames, are shown in (b), and their correlation is shown in (c).

## 3. Results

### 3.1. Atmospheric coherent structure characterization

The image classification process shows that vortical coherent structures appear frequently at an elevation where they can interact with wind turbines (Figure 4a). Furthermore, these coherent structures are highly intermittent, with the amount depending on wind speed (Figure 4b). In the Apr 2018 dataset with a mean wind speed of $\bar{U}_\infty = 2.6$ m/s, coherent structures occur 3% of the time (percent of frames labelled with 1 by the classifier). Both Feb 2019 datasets have higher mean wind speeds of $\bar{U}_\infty = 7.0$ m/s for Feb 2019a and $\bar{U}_\infty = 11.4$ m/s Feb 2019b, corresponding to rates of coherent structure appearance of 63% and 61%, respectively. Figure 4c shows the relationship between the level of coherent structures and the instantaneous wind speed. A sharp increase in the appearance of coherent structures is observed around wind speeds of 4 m/s. These findings are comparable to those presented by Träumner et al. (2015), who also observed significantly fewer atmospheric coherent structures at wind speeds below 4 m/s. Note that the level of coherent structures does not necessarily correspond to the turbulence intensity of the inflow calculated using the nacelle anemometer. The average turbulence intensity of all periods with coherent structures observed is 0.16, very close to that of the periods without structures (0.15).

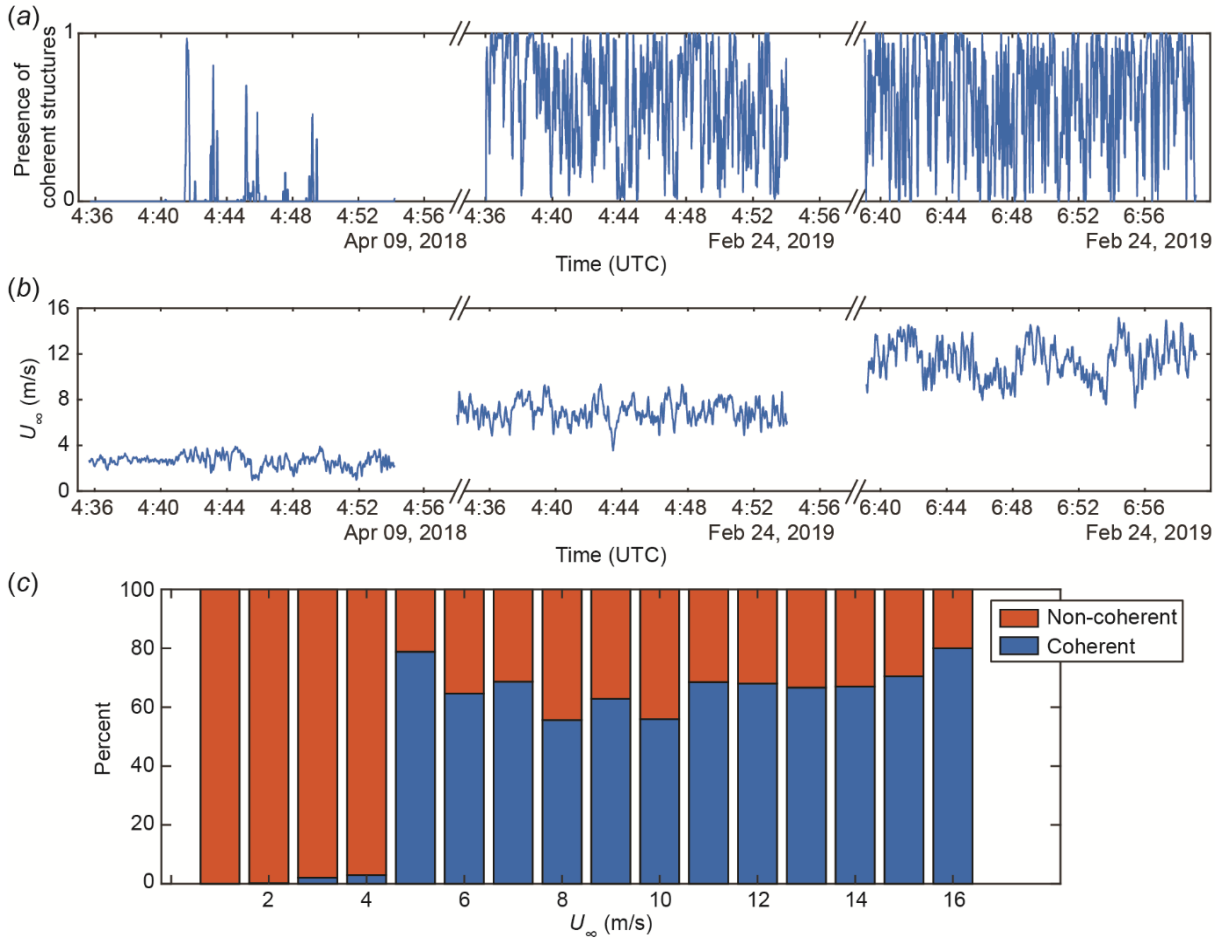

Fig. 4. Relationship between wind speed and the level of atmospheric vortical coherent structures observed, with (a) a time series of the level of coherent structures, as determined by the image classification method, for all three datasets (smoothed with a timescale of 3 s to facilitate visualization) and (b) a time series of wind speed at turbine hub height for all three datasets. (c) Percentage of frames labelled as coherent or non-coherent by the image classifier for each instantaneous wind speed. Note that non-coherent indicates a lack of *vortical* coherent structures. Other types of structures may be present.

In addition to characterizing the intermittency of atmospheric coherent structures, we also quantify two different structure scales which we term the "packet length" and the "vortex size". The packet length approximates the streamwise length of a group of vortical coherent structures in the ABL at the elevation of the region of interest shown in Figure 2, i.e., 35 m to 80 m. This scale is calculated using the amount of time that coherent structures are consistently in the inflow, determined by the time the coherent structure classification label stays above 0.5 (Figure 5a). The time scale is converted to length using the mean wind speed of each dataset, per Taylor's frozen turbulence hypothesis (Figure 5b). Under the conditions presented here, coherent structure packets can extend beyond 400 m, with an increase in the number of long packets at higher wind speeds. Figure 5c shows the probability distribution function (PDF) of the packet length scale, which peaks at the smallest detectable packet scale, and decreases monotonically with increasing packet length. These results are consistent with the findings of Lee & Sung (2011), who investigated very-large-scale coherent structures in a canonical turbulent boundary layer using direct numerical simulation, and Ganapathisubramani, Longmire, & Marusic (2003), who observed hairpin vortex packets extending up to twice the boundary layer thickness in length. The long tail of the distribution indicates the presence of analogous long packets in the ABL. Previous studies have observed these structures extending up to 1500 m in the streamwise direction (Träumner et al., 2015). Such packet lengths are not observed here, though this discrepancy is likely caused by out-of-plane motions that would make very long structures appear as multiple separate structures. The organization of vortical coherent structures into packets and the shape of the packet length distribution suggest the structures observed here are signatures of canonical turbulent boundary layer structures, which originate from a disturbance at the wall and grow outwards (Adrian, Meinhart, & Tomkins, 2000). As we will show in Section 3.2, the packet length scale is also an important factor in determining the impact of coherent structures on turbine structural loading.

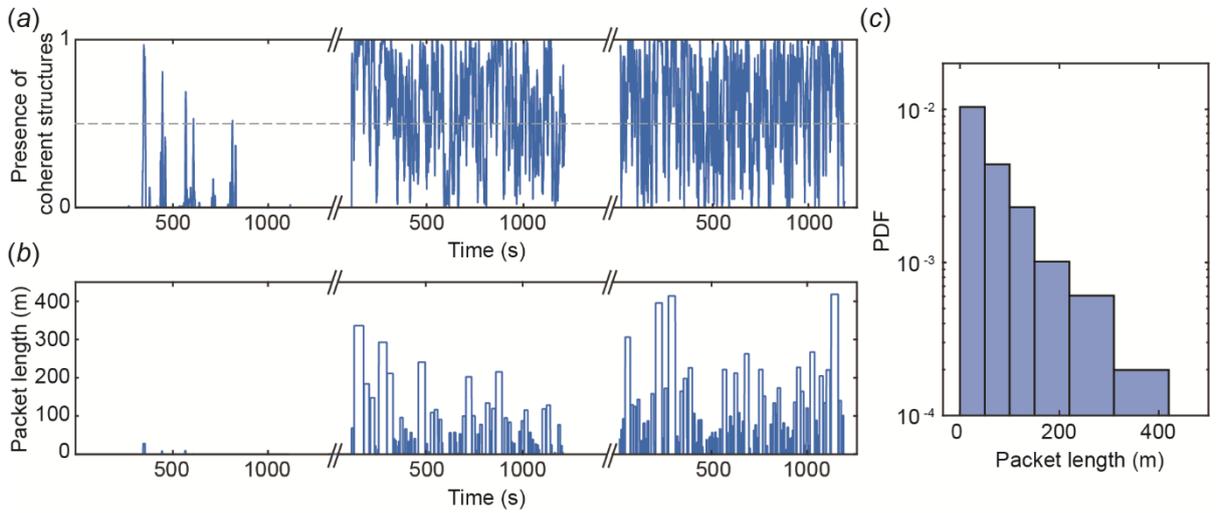

Fig. 5. (a) The level of coherent structures from image classification, with a dashed grey line indicating the threshold, used to determine (b) the coherent structure packet length scale. (c) Probability distribution function (PDF) of the packet length for all three datasets. Note that the bin size is gradually increased for events with larger packet lengths to ensure sufficient statistical convergence for events with lower probability of occurrence.

The vortex size is estimated based on the void properties in the atmospheric flow images. Note that these results once again represent an approximation of vortex size, as only a cross-section of the three-dimensional structure is captured in the FOV. The voids are extracted from the inflow region of the images using a combination of image intensity thresholding and edge detection. This image processing method is described in detail in Abraham & Hong (2020). The cross-sectional area ($A$) of the largest atmospheric coherent structure in each frame is determined by the number of pixels, and an equivalent diameter of the structure is calculated using $d_{eq} = \sqrt{4A/\pi}$ (Figure 6a). The largest void in each frame is used because larger structures are expected to have a stronger impact on the turbine than smaller structures. Because of the centrifugal effect of the fluid rotation on the snow particles, the edges of the voids are determined by the Stokes number, $St = \tau_p/\tau_f$, where $\tau_p$ is the particle time scale and $\tau_f$ is the flow time scale (Eaton & Fessler, 1994). As $\tau_f$ is determined by the strength (i.e., circulation, $\Gamma$) of the vortex causing the snow particle void, the diameter of the void is directly related to the vortex strength. Therefore, the void boundaries represent a circulation threshold that is approximately equal to that determined by Hong et al. (2014), i.e., $\Gamma \approx 6$ m$^2$/s. Here, we use $d_{eq}$ to investigate the vortex size for all three datasets (Figure 6b). As evidenced by the size distributions shown in Figure 6c, the Apr 2018 dataset exhibits the fewest and smallest vortices, with $\overline{d_{eq}} = 1.0$ m, where the overbar denotes the mean. The vortices in the Feb 2019a dataset have $\overline{d_{eq}} = 1.5$ m, and Feb 2019b includes the most vortices with the largest size ($\overline{d_{eq}} = 2.4$ m). The vortex size discrepancy between the three datasets is attributed to their different values of friction velocity, $U_\tau$. Previous studies have shown that the vortex diameter increases with distance from the wall, normalized by the viscous length scale, i.e., $z^+ = zU_\tau/\nu$ (Robinson, 1990). Therefore, though all three FOVs are located at the same elevation, they capture different parts of the boundary layer, with larger vortices occurring at larger values of $z^+$ (Figure 6d). These findings are also consistent with the results of Ganapathisubramani et al. (2003) for a canonical turbulent boundary layer. They showed that the minimum vortex packet length, which corresponds to a single hairpin vortex, increases with increasing values of $z^+$. Note that a digital inline holography sensor was used to measure the snowflake size (see Nemes et al., 2017 for a detailed description of this method), and the mean snowflake equivalent diameter was consistently between 0.3-0.4 mm for all three datasets. Additionally, temperature and humidity conditions for all three datasets were within 1°C and 1%, respectively, leading the snowflakes of the same shape (Pruppacher & Klett, 2010). Therefore, differences in coherent structure size cannot be attributed to discrepancies in snowflake properties. As we will show in Section 3.3, vortex size is also related to the impact of inflow coherent structures on wind turbine power production and wake behaviour.

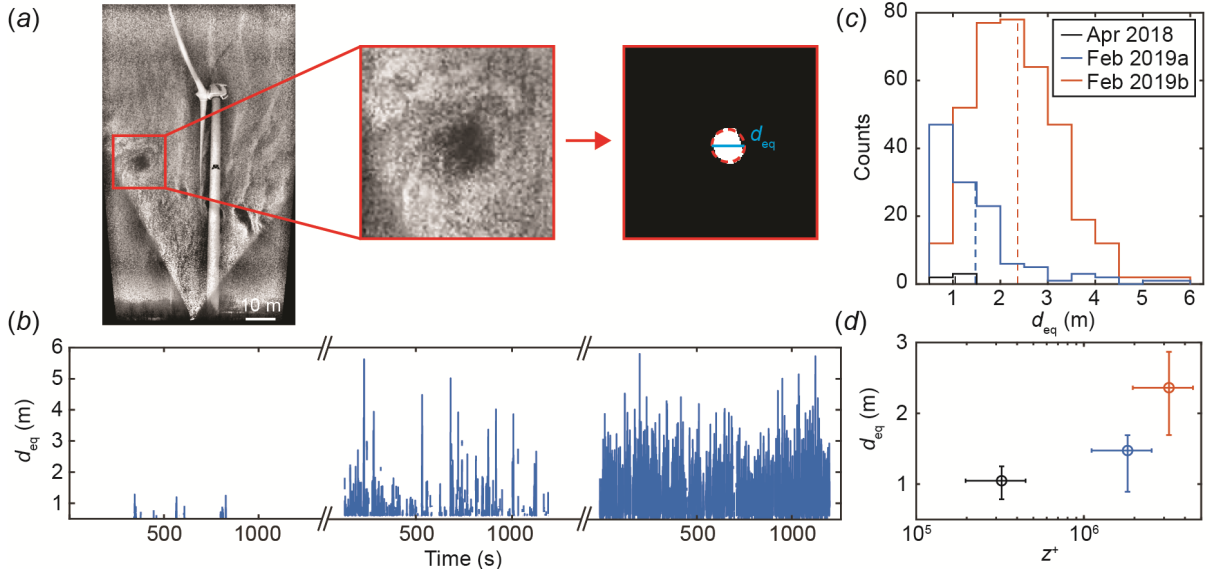

Fig. 6. Vortex size characterization, including (a) a sample image with a coherent structure in the inflow, and the resulting extracted void. The equivalent diameter ($d_{eq}$) of the void is indicated by the blue line bisecting the red dashed circle. (b) A time series of $d_{eq}$ over all three datasets. (c) Histogram of $d_{eq}$ for each dataset, with dashed vertical lines representing the mean. (d) Relationship between $z^+$ and $d_{eq}$, with the circles indicating the mean values and the error bars representing the standard deviations. A log scale is used for $z^+$, as ABL quantities typically vary with the log of the elevation.

### 3.2. Impact on structural loading

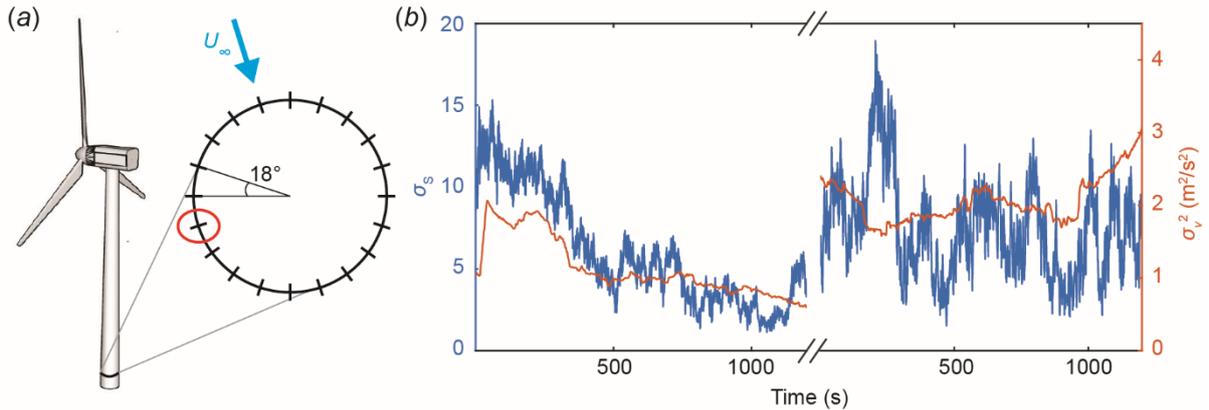

Fig. 7. (a) Schematic showing the strain gauges located around the base of the Eolos turbine tower. The strain gauge used for the following analysis is circled in red and the wind direction is indicated by a blue arrow. (b) Time series of the standard deviation of strain ($\sigma_s$) and the square of the standard deviation of the spanwise wind component ($\sigma_v^2$) for both Feb 2019 datasets. Note that the Apr 2018 dataset has been removed, as the turbine is not producing power for 97% of the recorded period.

We now investigate the impact of inflow coherent structures on the structural loading of the utility-scale wind turbine. As mentioned in Section 2.1, 20 strain gauges are mounted around the base of the Eolos turbine support tower. In the current study, we focus on the lateral strain, i.e., the strain gauge located perpendicular to the wind direction, as we observed the strongest relationship between inflow turbulence and strain in this direction. Because of lateral symmetry, all analysis is conducted on a single strain gauge located 90° anticlockwise from the incoming wind for both Feb 2019 datasets – Apr 2018 is removed from the analysis, as the turbine is not producing power for 97% of the recorded period, and the remaining data is insufficient to derive meaningful conclusions with statistical significance (Figure 7a). The analysis focuses on the standard deviation of lateral strain ($\sigma_s$), as strain fluctuations indicate fatigue loading on the turbine. In Figure 7b, $\sigma_s$ is compared to $\sigma_v^2$, the square of the standard deviation of spanwise velocity, showing they follow similar large-scale trends. Both $\sigma_s$ and $\sigma_v^2$ are calculated over a 7.5-min moving window. This window length was chosen as it yields the maximum correlation between $\sigma_s$ and $\sigma_v^2$, suggesting it is the timescale at which their interaction occurs. Based on this relationship between $\sigma_s$ and $\sigma_v^2$, fluctuations in the drag force exerted by the spanwise component of the wind cause fluctuating loads on the tower. Indeed, the drag force of a fluid on a cantilevered beam (like the turbine support tower) is proportional to the square of the velocity and directly proportional to the strain at the base of the beam.

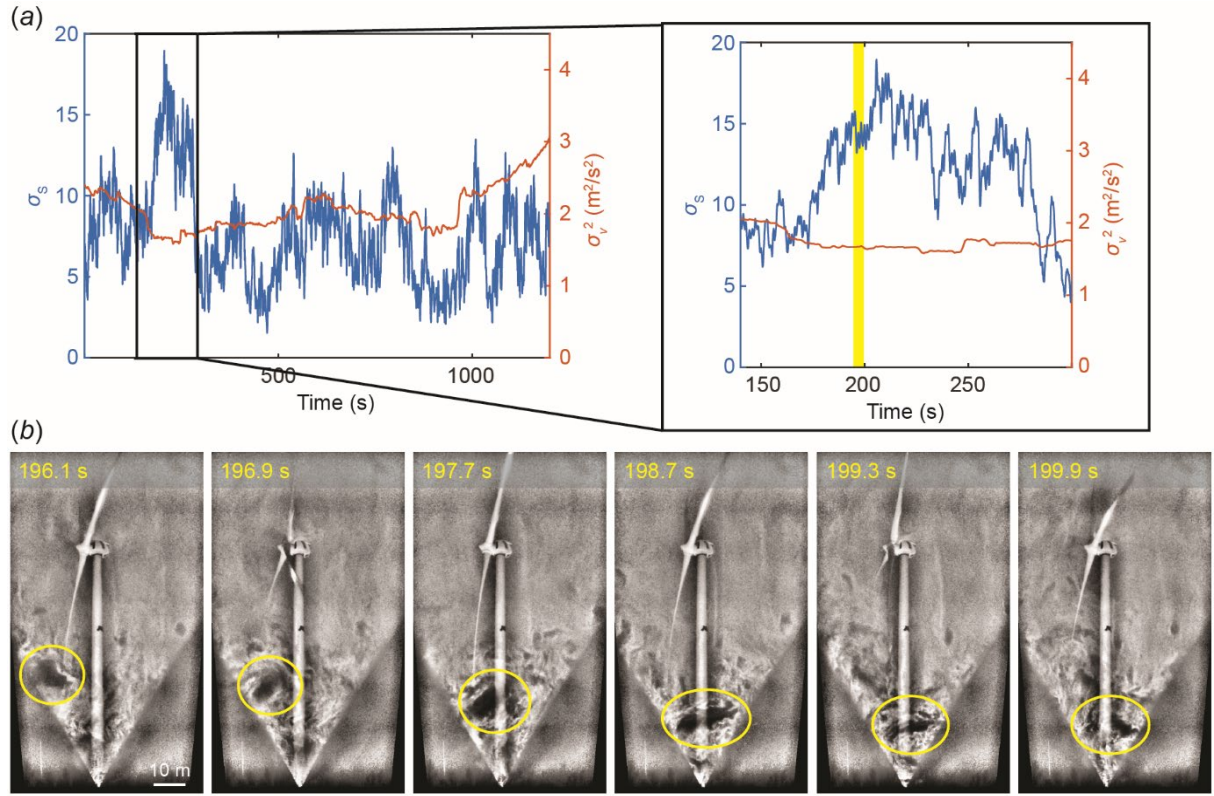

However, some periods are observed where $\sigma_s$ does not follow $\sigma_v^2$. One example of this deviation is exhibited in Figure 8a, where a strong increase in $\sigma_s$ is observed, while $\sigma_v^2$ stays relatively constant. In Figure 8b, a proposed explanation for this increase is demonstrated with several flow visualization images from this time period. These images clearly show a strong atmospheric coherent structure entering the FOV and interacting with the turbine tower, which induces fluctuating loads on the tower. Several such coherent structures are observed during this period of increasing $\sigma_s$, though only the clearest is shown here. This example suggests that coherent structures generate additional structural loading on the turbine beyond that induced by velocity fluctuations, consistent with the findings of Kelley et al. (2005).

Our results further show that tower strain fluctuations increase with increasing coherent structure packet length (Figure 9). Once again, the Apr 2018 dataset is not included in the analysis due to the limited amount of data where the turbine is operating. However, both Feb 2019 datasets clearly exhibit the presence of long packets coinciding with peaks in $\sigma_s$ (Figure 9a). Binning $\sigma_s$ by packet length shows that the general positive relationship between packet length and $\sigma_s$ is consistent throughout the entirety of the data collected (Figure 9b). Furthermore, the correlation between the two variables is 0.6 with $p \ll 0.01$. These findings are consistent with the understanding developed in the study of lower Reynolds-number turbulent boundary layers that longer packets contribute significantly more turbulence production and momentum transport than shorter packets (Ganapathisubramani et al., 2003). In interacting with a utility-scale wind turbine, this additional turbulence and momentum leads to increased fluctuating loads on the structure. This phenomenon has not been previously observed at such high Reynolds numbers as those investigated in the current study (Re~$10^7$). This finding also has important implications for wind farm siting decisions, as landscape features or buildings that generate large coherent structures can lead to additional structural fatigue loading.

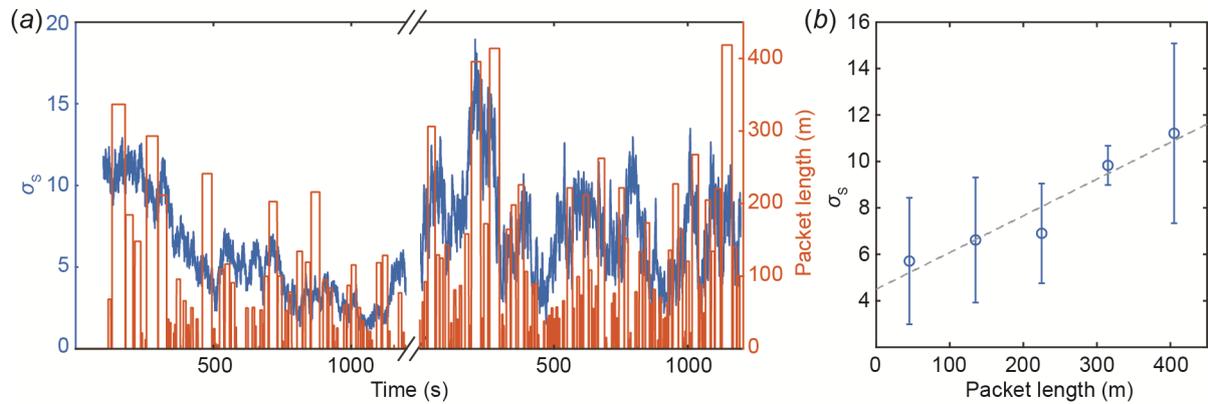

Fig. 9. Relationship between coherent structure packet length and standard deviation of lateral tower strain ($\sigma_s$), including (a) a time series across both Feb 2019 datasets, and (b) a scatter plot with the $\sigma_s$ data points binned by packet length. The circles indicate the mean value of $\sigma_s$ for each value of packet length, and the error bars indicate the standard deviation. Once again, the Apr 2018 dataset has been removed due to the limited duration of turbine operation (3%).

### 3.3. Impact on power production and wake behaviour

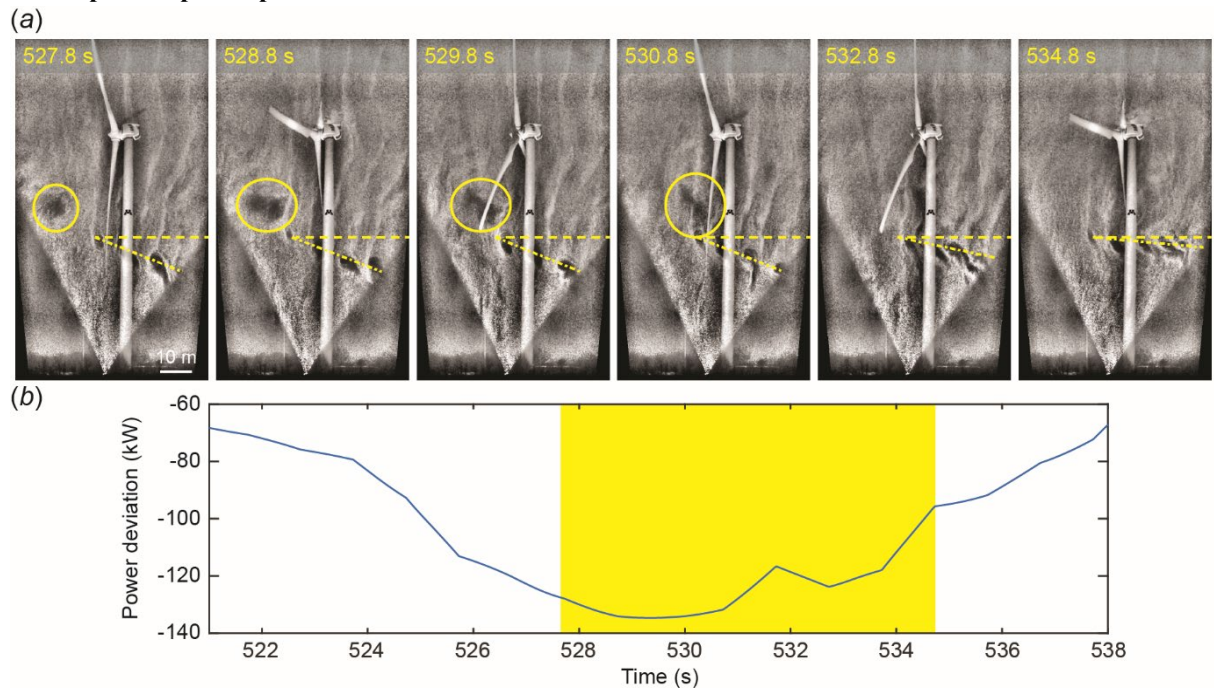

Fig. 10. Example of a coherent structure interacting with the turbine and inducing (a) a reduction in wake expansion and (b) a reduction in power generation. In (a), the coherent structure is circled in yellow, and the wake expansion angle is indicated by the yellow lines. The dashed line represents the bottom blade tip elevation, and the dot-dashed line shows the position of the tip vortices. The yellow region marked in (b) indicates the time period shown in the images. Power deviation is defined as the expected power production at the given wind speed based on the power curve, subtracted from the actual power produced.

We next explore the effect of coherent structures on turbine power production and wake behaviour. When the turbine is producing power and extracting energy from the wind, the wake typically expands as the air slows down due to mass and momentum conservation. Previous studies have observed some periods of wake contraction caused by changes in the pitch of the turbine blades (Abraham & Hong, 2020; Dasari et al., 2019). These blade pitch changes occur when the turbine is operating above the rated wind speed and the turbine reduces the angle of attack of the blades to regulate loading on the turbine. In the current study, we focus on periods when the turbine is operating below the rated wind speed and the blade pitch is not changing. Instances of wake contraction are also observed under these conditions, suggesting the existence of an additional mechanism leading to wake contraction. Closer investigation reveals that these contraction periods occur when there are coherent structures in the inflow that interact with the turbine (Figure 10a). Furthermore, they coincide with a reduction in power generation compared to the expected performance based on the turbine power curve (Figure 10b).

To characterize this wake contraction behaviour, we develop a method to quantify the wake expansion angle. Using an image processing technique similar to that described in Section 3.1, binary images of the tip vortices are extracted from the enhanced snow images. The centroid of the tip vortex nearest the turbine is determined from the binary image, and the angle of the centroid from the elevation of the bottom turbine blade tip ($\varphi_\mathrm{w}$) is calculated (Figure 11). Only the first tip vortex downstream of the turbine tower is used to calculate $\varphi_\mathrm{w}$ in order to minimize distortion caused by interactions between adjacent vortices as they advect downstream. The wake expansion angle is quantified for the entire Feb 2019a dataset, where the turbine operates below the rated wind speed. Any periods of wake contraction observed while the turbine is operating under these conditions cannot be attributed to changes in blade pitch, which is fixed to a minimal value of 1° to maximize power generation.

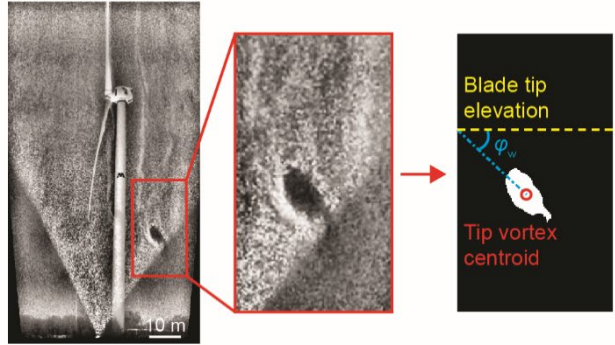

Fig. 11. Definition of wake expansion angle, $\varphi_\mathrm{w}$, as determined using the centroid of the bottom blade tip vortex and the elevation of the tip of the bottom turbine blade.

Figure 12 shows a time series of $\varphi_\mathrm{w}$, power deviation, and vortex size ($d_\mathrm{eq}$, defined in Section 3.1). Power deviation is defined as expected power subtracted from actual power, where expected power is determined using the Eolos turbine power curve and the instantaneous hub-height wind speed. The expected power is filtered and shifted in time to maximize the correlation with the actual power. A negative value of power deviation indicates the turbine is under-producing, while a positive value indicates the turbine is producing more power than expected. Five periods are highlighted in Figure 12 where a large power deviation is observed. For the three periods where this power deviation is negative (marked in blue), $\varphi_\mathrm{w}$ is also negative, indicating wake contraction. All five periods also coincide with above-average values of $d_\mathrm{eq}$. The images showing the spanwise vorticity ($\omega_y$) and vector fields superimposed on the snow visualization images for each period provide further insight into the cause of the power surpluses and deficits observed. For the power deficit periods, large coherent structures with negative vorticity are observed in the inflow. For the power surplus periods (marked in yellow), large structures with positive vorticity are detected. We attribute this dependence on rotation direction to lift generation on the turbine blades. Lift is directly proportional to the circulation around the blade cross-sections per the Kutta-Joukowsky equation (Sherwood, 1946). The positive vorticity of the vortices shed from the blade tips indicates that the bound circulation on the blades is also positive. Therefore, negative vorticity in the inflow neutralizes some of the lift, reducing the power generation and inducing wake contraction, while positive vorticity enhances the lift, increasing power generation.

Not every period with a large power surplus or deficit demonstrates this clear relationship with inflow coherent structure vorticity because of the complex nature of field data. In the field, turbine and wake behaviour are influenced by many different variables, making it difficult to establish a clear one-to-one correspondence between variables. In particular, the FOV of the flow visualization data only captures a cross-section of the inflow, so the three-dimensionality or structures appearing outside of the visualized plane cannot be detected. Some of the large power deviations are likely caused by structures outside of the FOV plane. Still, a statistically robust relationship between power deviation and vortex size is observed (Figure 13). The histogram of power deviation magnitude conditionally sampled by inflow vortex size shows a clear separation between vortices that are below and above the mean value of $d_\mathrm{eq}$. This separation is further strengthened when comparing vortices from the bottom and top quartile of $d_\mathrm{eq}$, which correspond to the smallest and largest quarter, respectively, of all vortices detected in the inflow while the turbine is operating below the rated wind speed. These results are confirmed to be statistically significant using a two-sample Kolmogorov-Smirnov test ($p \ll 0.01$).

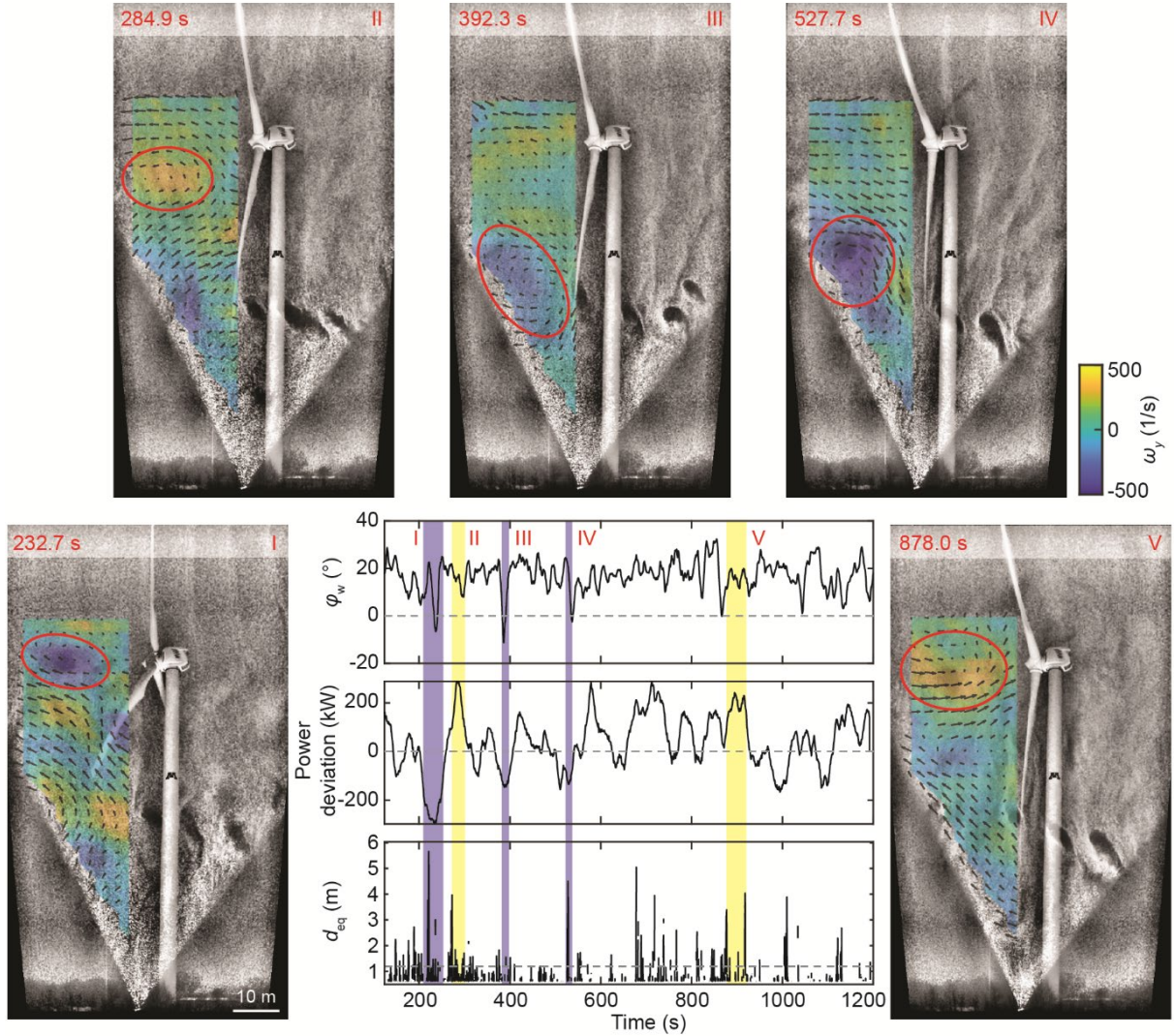

Fig 12. Time series of wake expansion angle ($\varphi_w$), power deviation (expected power subtracted from actual power), and vortex size ($d_{eq}$). The blue and yellow bars indicate periods with strong power deficits and surpluses, respectively. The red numbers correspond to the images show around the plot, which are sample enhanced snow particle images superimposed with spanwise vorticity and vector fields from each highlighted time period. The vector fields represent the fluid velocity with the mean subtracted. Note that the region of positive vorticity around the bottom blade tip in some images is caused by recently generated tip vortices. Supplemental videos 1-5 show the vorticity and vector fields for the duration of each of the five highlighted periods.

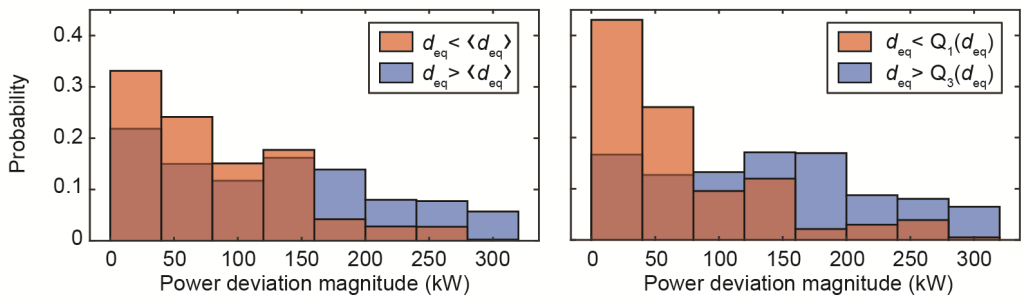

Fig. 13. Magnitude of power deviation magnitude conditionally sampled by vortex size. The plot on the left compares the vortices below and above the mean value of vortex size ($\langle d_{eq} \rangle$), while the plot on the right compares vortices in the bottom and top quartile of vortex size ($Q_1(d_{eq})$ and $Q_3(d_{eq})$), respectively.

## 4. Conclusion

In this study, we characterize turbulent coherent structures in the atmospheric boundary layer and investigate their impact on utility-scale wind turbine loading, power generation, and wake behaviour. This investigation is conducted using flow visualization with natural snowfall with a field of view spanning the inflow and near wake of a 2.5 MW turbine. Coherent vortical structures in the inflow are detected using a manually trained image classifier and the flow field is quantified with super-large-scale particle image velocimetry. This method facilitates the visualization and analysis of atmospheric structures that could not previously be observed due to limitations in the spatio-temporal resolution of conventional field-scale measurement techniques. Three datasets with different conditions reveal the substantial increase in inflow coherent structures with increasing wind speed. The coherent structure packet length, determined by the duration with coherent structures consistently in the inflow, exhibits a long-tailed distribution similar to that observed for hairpin vortex packets in canonical turbulent boundary layers. Meanwhile, the inflow vortex size distribution varies between the three datasets due to differences in the friction velocity. Turbine structural loading is shown to depend on spanwise wind speed fluctuations, though some periods exhibit disproportionately high tower strain fluctuations due to interaction with coherent structures. Tower strain fluctuations also increase significantly with increases in coherent structure packet length. Finally, we observe a relationship between large vortical coherent structures in the inflow and deviations in turbine power production from the expected value based on the power curve. These deviations are attributed to rotation direction of the atmospheric structures, with positive vorticity leading to a power surplus, and negative vorticity leading to a power deficit along with wake contraction.

These findings have implications for wind turbine design and siting decisions. On the one hand, coherent structures tend to induce increased structural loading and deviations from expected turbine performance, both of which increase the cost and uncertainty of wind farm operation. On the other hand, our findings suggest that coherent structures with the same rotation direction as the blade circulation have the potential to boost power generation on short timescales. Consequently, the size, rotation direction, and frequency of occurrence of vortical coherent structures should be evaluated alongside mean wind conditions at prospective wind farm sites. These factors should then be included in considerations for turbine design and layout decisions. These findings will be particularly important in situations when obstacles such as buildings or tree are located upstream of a turbine, as they will generate additional coherent structures. However, vortical coherent structures are present even in flat terrain such as that investigated in the current study. In this case, because of the direction of shear in the ABL, large structures with negative vorticity are more common than those with positive vorticity (Heisel et al., 2018a). Consequently, it may be beneficial to select turbine rotation direction such that blade circulation is negative as well. Finally, we reveal an additional mechanism leading to turbine wake contraction. Dasari et al. (2019) showed that wake contraction occurs when the turbine is operating above the rated wind speed and the blade pitch changes, deflecting the turbine into its own wake. In the current study, we further show that wake contraction can also be induced in below rated operation by coherent vortical structures in the inflow reducing the lift generated by the blades. This near-wake behaviour can significantly impact wake development downstream, including wake length, recovery, and mixing.

One limitation of this study has already been mentioned, i.e., a single plane of visualization data which does not allow us to observe the entire three-dimensional shape of the coherent structures or detect structures outside of the plane. The current study also focuses exclusively on vortical coherent structures, as they can be easily visualised using natural snowfall. However, the impact of other boundary layer coherent structures (e.g., streaks, rolls) has not been evaluated. Most of the coherent structures observed in the current study occurred under near-neutral atmospheric stability conditions. Previous studies have shown that stability affects the properties of large-scale regions of coherent velocity fluctuations (Barthlott et al., 2007), so we expect the vortical atmospheric structures investigated here to exhibit similar dependencies. In addition, all three datasets were collected at the same site, which has a relatively uniform surface roughness. Surface roughness is expected to affect the properties of the coherent structures appearing in the ABL, but we do not expect the main trends presented here to be significantly modified. Finally, the turbine geometry (e.g., rotor diameter, height, blade profile, etc.) will influence the magnitude of the coherent structure impact on loading, power, and wake. However, the physical mechanisms described here will remain relevant regardless of turbine design.

Acknowledgments. We thank the students and engineers from St. Anthony Falls Laboratory for assistance with the experiments, including S Riley, B Li, T Dasari, Y Wu, C Li, K Lim, Z Zhou, J Tucker, C Milliren, J Marr, R Christopher, M Lueker, B Erickson, and C Feist. We also thank A Kannan for assistance with preliminary data analysis.

Funding statement. This work was supported by the National Science Foundation CAREER award (NSF-CBET-1454259), Xcel Energy through the Renewable Development Fund (grant RD4-13) as well as IonE of University of Minnesota.

Competing interests. None

Data availability statement. Data and codes are available upon reasonable request from the authors.

Ethical standards. The research meets all ethical guidelines, including adherence to the legal requirements of the study country.

Author contributions. Conceptualization: A.A; J.H. Funding acquisition: J.H. Methodology: A.A; J.H. Data curation: A.A. Software: A.A. Investigation: A.A.; J.H. Data visualisation: A.A. Supervision: J.H. Writing original draft: A.A; J.H. All authors approved the final submitted draft.

**References**

Abraham, A., Dasari, T., & Hong, J. (2019). Effect of turbine nacelle and tower on the near wake of a utility-scale wind turbine. *Journal of Wind Engineering and Industrial Aerodynamics*, *193*, 103981. https://doi.org/10.1016/j.jweia.2019.103981

Abraham, A., & Hong, J. (2020). Dynamic wake modulation induced by utility-scale wind turbine operation. *Applied Energy*, *257*, 114003. https://doi.org/10.1016/j.apenergy.2019.114003

Abraham, A., & Hong, J. (2021). Operational-dependent wind turbine wake impact on surface momentum flux. *Renewable and Sustainable Energy Reviews*, *144*(July 2020), 111021. https://doi.org/10.1016/j.rser.2021.111021

Adrian, R. J., Meinhart, C. D., & Tomkins, C. D. (2000). Vortex organization in the outer region of the turbulent boundary layer. *Journal of Fluid Mechanics*, *422*, 1–54. https://doi.org/10.1017/S0022112000001580

Adrian, Ronald J. (2007). Hairpin vortex organization in wall turbulence. *Physics of Fluids*, *19*, 041301. https://doi.org/10.1063/1.2717527

Alcayaga, L., Larsen, G. C., Kelly, M., & Mann, J. (2020). Large-scale coherent structures in the atmosphere over a flat terrain. *Journal of Physics: Conference Series*, *1618*, 062030. https://doi.org/10.1088/1742-6596/1618/6/062030

Barthlott, C., Drobinski, P., Fesquet, C., Dubos, T., & Pietras, C. (2007). Long-term study of coherent structures in the atmospheric surface layer. *Boundary-Layer Meteorology*, *125*(1), 1–24. https://doi.org/10.1007/s10546-007-9190-9

Chamorro, L. P., Hill, C., Neary, V. S., Gunawan, B., Arndt, R. E. A., & Sotiropoulos, F. (2015). Effects of energetic coherent motions on the power and wake of an axial-flow turbine. *Physics of Fluids*, *27*(5). https://doi.org/10.1063/1.4921264

Chamorro, L. P., Lee, S. J., Olsen, D., Milliren, C., Marr, J., Arndt, R. E. A., & Sotiropoulos, F. (2015). Turbulence effects on a full-scale 2.5 MW horizontal-axis wind turbine under neutrally stratified conditions. *Wind Energy*, *18*, 339–349. https://doi.org/10.1002/we.1700

Cheliotis, I., Dieudonné, E., Delbarre, H., Sokolov, A., Dmitriev, E., Augustin, P., & Fourmentin, M. (2020). Detecting turbulent structures on single Doppler lidar large datasets: an automated classification method for horizontal scans. *Atmospheric Measurement Techniques*, *13*, 6579–6592. https://doi.org/10.5194/amt-2020-82

Dasari, T., Wu, Y., Liu, Y., & Hong, J. (2019). Near-wake behaviour of a utility-scale wind turbine. *Journal of Fluid Mechanics*, *859*, 204–246. https://doi.org/10.1017/jfm.2018.779

Eaton, J. K., & Fessler, J. R. (1994). Preferential concentration of particles by turbulence. *International Journal of Multiphase Flow*, *20*, 169–209.

Frandsen, S. T. (2007). *Turbulence and turbulence- generated structural loading in wind turbine clusters*. *Risoe National Laboratory. Risoe-R; No. 1188* (Vol. 1188). https://doi.org/Riso-R-1188

Ganapathisubramani, B., Longmire, E. K., & Marusic, I. (2003). Characteristics of vortex packets in turbulent boundary layers. *Journal of Fluid Mechanics*, *478*(478), 35–46. https://doi.org/10.1017/S0022112002003270

Heisel, M., Dasari, T., Liu, Y., Hong, J., Coletti, F., & Guala, M. (2018). The spatial structure of the logarithmic region in very-high-Reynolds-number rough wall turbulent boundary layers. *Journal of Fluid Mechanics*, *857*, 704–747. https://doi.org/10.1017/jfm.2018.759

Heisel, M., De Silva, C. M., Hutchins, N., Marusic, I., & Guala, M. (2020). On the mixing length eddies and logarithmic mean velocity profile in wall turbulence. *Journal of Fluid Mechanics*, *887*, 1–13. https://doi.org/10.1017/jfm.2020.23

Heisel, M., Hong, J., & Guala, M. (2018). The spectral signature of wind turbine wake meandering: A wind tunnel and field-scale study. *Wind Energy*, *21*(9), 715–731. https://doi.org/10.1002/we.2189

Hommema, S. E., & Adrian, R. J. (2003). Packet structure of surface eddies in the atmospheric boundary layer. *Boundary-Layer Meteorology*, *106*(1), 147–170. https://doi.org/10.1023/A:1020868132429

Hong, J., & Abraham, A. (2020). Snow-powered research on utility-scale wind turbine flows. *Acta Mechanica Sinica*, *36*(2), 339–355. https://doi.org/10.1007/s10409-020-00934-7

Hong, J., Toloui, M., Chamorro, L. P., Guala, M., Howard, K., Riley, S., … Sotiropoulos, F. (2014). Natural snowfall reveals large-scale flow structures in the wake of a 2.5-MW wind turbine. *Nature Communications*, *5*, 4216. https://doi.org/10.1038/ncomms5216

Karagali, I., Mann, J., Dellwik, E., & Vasiljević, N. (2018). New European Wind Atlas: The Osterild balconies

experiment. *Journal of Physics: Conference Series*, *1037*, 052029. https://doi.org/10.1088/1742-6596/1037/5/052029

Kelley, N. D., Jonkman, B. J., Scott, G. N., Bialasiewicz, J. T., & Redmond, L. S. (2005). The Impact of Coherent Turbulence on Wind Turbine Aeroelastic Response and Its Simulation. In *WindPower 2005* (p. 22). Retrieved from http://www.nrel.gov/docs/fy05osti/38074.pdf

Lee, J. C. Y., & Fields, M. J. (2021). An overview of wind-energy-production prediction bias, losses, and uncertainties. *Wind Energy Science*, *6*(2), 311–365. https://doi.org/10.5194/wes-6-311-2021

Lee, J. H., & Sung, H. J. (2011). Very-large-scale motions in a turbulent boundary layer. *Journal of Fluid Mechanics*, *673*(April 2011), 80–120. https://doi.org/10.1017/S002211201000621X

Li, C., Abraham, A., Li, B., & Hong, J. (2020). Incoming flow measurements of a utility-scale wind turbine using super-large-scale particle image velocimetry. *Journal of Wind Engineering and Industrial Aerodynamics*, *197*, 104074. https://doi.org/10.1016/j.jweia.2019.104074

Li, D., & Bou-Zeid, E. (2011). Coherent structures and the dissimilarity of turbulent transport of momentum and scalars in the unstable Atmospheric surface layer. *Boundary-Layer Meteorology*, *140*(2), 243–262. https://doi.org/10.1007/s10546-011-9613-5

MathWorks. (2020). Image Category Classification Using Bag of Features. Retrieved April 14, 2020, from https://www.mathworks.com/help/vision/ug/image-category-classification-using-bag-of-features.html;jsessionid=c7f46355be1453ca17707c1ef460

Medici, D., & Alfredsson, P. H. (2006). Measurements on a wind turbine wake: 3D effects and bluff body vortex shedding. *Wind Energy*, *9*(3), 219–236. https://doi.org/10.1002/we.156

Nemes, A., Dasari, T., Hong, J., Guala, M., & Coletti, F. (2017). Snowflakes in the atmospheric surface layer: observation of particle-turbulence dynamics. *Journal of Fluid Mechanics*, *814*, 592–613. https://doi.org/10.1017/jfm.2017.13

Oncley, S. P., Hartogensisa, O., & Tong, C. (2016). Whirlwinds and hairpins in the atmospheric surface layer. *Journal of the Atmospheric Sciences*, *73*(12), 4927–4943. https://doi.org/10.1175/JAS-D-15-0368.1

Park, J., Basu, S., & Manuel, L. (2014). Large-eddy simulation of stable boundary layer turbulence and estimation of associated wind turbine loads. *Wind Energy*, *17*, 359–384. https://doi.org/10.1002/we.1580

Pruppacher, H. R., & Klett, J. D. (2010). *Microphysics of Clouds and Precipitation* (2nd ed.). Springer, Dordrecht.

Robinson, S. K. (1990). A Review of Vortex Structures and Associated Coherent Motions in Turbulent Boundary Layers. In A. Gyr (Ed.), *Structure of Turbulence and Drag Reduction* (pp. 23–50). Zurich: Springer-Verlag Berlin Heidelberg. https://doi.org/10.1007/978-3-642-50971-1_2

Sherwood, A. W. (1946). *Aerodynamics*. New York: McGraw-Hill.

Stull, R. B. (1988). *An introduction to boundary layer meteorology*. Dordrecht: Kluewer Academic Publishers.

Thielicke, W., & Stamhuis, E. J. (2014). PIVlab – Towards User-friendly, Affordable and Accurate Digital Particle Image Velocimetry in MATLAB. *Journal of Open Research Software*, *2*. https://doi.org/10.5334/jors.bl

Toloui, M., Riley, S., Hong, J., Howard, K., Chamorro, L. P., Guala, M., & Tucker, J. (2014). Measurement of atmospheric boundary layer based on super-large-scale particle image velocimetry using natural snowfall. *Experiments in Fluids*, *55*, 1737. https://doi.org/10.1007/s00348-014-1737-1

Träumner, K., Damian, T., Stawiarski, C., & Wieser, A. (2015). Turbulent Structures and Coherence in the Atmospheric Surface Layer. *Boundary-Layer Meteorology*, *154*(1), 1–25. https://doi.org/10.1007/s10546-014-9967-6

Veers, P., Dykes, K., Lantz, E., Barth, S., Bottasso, C. L., Carlson, O., … Wiser, R. (2019). Grand challenges in the science of wind energy. *Science*, *366*, eaau2027. https://doi.org/10.1126/science.aau2027

Wu, Y., & Christensen, K. T. (2006). Population trends of spanwise vortices in wall turbulence. *Journal of Fluid Mechanics*, *568*(1952), 55–76. https://doi.org/10.1017/S002211200600259X

Wu, Yu-ting, & Porte-Agel, F. (2012). Atmospheric Turbulence Effects on Wind-Turbine Wakes: An LES Study. *Energies*, *5*, 5340–5362. https://doi.org/10.3390/en5125340